# Super-resolved multispectral lensless microscopy via angle-tilted, wavelength-multiplexed ptychographic modulation

PENGMING SONG,[1,4] RUIHAI WANG,[2,4] JIAKAI ZHU,[2] TIANBO WANG,[2] ZICHAO BIAN,[2] ZIBANG ZHANG,[2] KAZUNORI HOSHINO,[2] MICHAEL MURPHY,[3] SHAOWEI JIANG,[2,*] CHENGFEI GUO,[2,*] AND GUOAN ZHENG[1,2]

[1]Department of Electrical and Computer Engineering, University of Connecticut, Storrs, CT, 06269, USA
[2]Department of Biomedical Engineering, University of Connecticut, Storrs, CT, 06269, USA
[3]Department of Dermatology, University of Connecticut Health Center, Farmington, Connecticut 06030, USA
[4]These authors contributed equally to this work
*Corresponding author: shaowei.jiang@uconn.edu or chengfei.guo@uconn.edu

Date: 05/19/2020

**We report an angle-tilted, wavelength-multiplexed ptychographic modulation approach for multispectral lensless on-chip microscopy. In this approach, we illuminate the specimen with lights at 5 wavelengths simultaneously. A prism is added at the illumination path for spectral dispersion. Lightwaves at different wavelengths, thus, hit the specimen at slightly different incident angles, breaking the ambiguities in mixed state ptychographic reconstruction. At the detection path, we place a thin diffuser in-between the specimen and the monochromatic image sensor for encoding the spectral information into 2D intensity measurements. By scanning the sample to different x-y positions, we acquire a sequence of monochromatic images for reconstructing the 5 complex object profiles at the 5 wavelengths. An up-sampling procedure is integrated into the recovery process to bypass the resolution limit imposed by the imager pixel size. We demonstrate a half-pitch resolution of 0.55 µm using an image sensor with 1.85-µm pixel size. We also demonstrate quantitative and high-quality multispectral reconstructions of stained tissue sections for digital pathology applications.**

*OCIS codes: (110.0180) Microscopy; (110.4234) Multispectral and hyperspectral imaging; (100.5070) Phase retrieval.*

Multispectral microscopy, in general, refers to the acquisition of 2D microscopic images at more than 3 wavelengths. Thanks to different spectral signatures of biochemical markers, multispectral microscopy can be used for better histopathological analysis of labeled tissue sections [1]. The adoption of multispectral microscopy in digital pathology allows pathologists to test new staining protocols, including multiplexed antibody studies, for better and faster diagnosis and prognosis of cancers and other diseases.

In lensless on-chip microscopy, object images are obtained using a cost-effective and compact system without any lens. Lensless color imaging at red, green, and blue wavelengths has been demonstrated using color image sensor, where color information is encoded via the RGB Bayer filter [2, 3]. Alternatively, one can sequentially illuminate the sample with different wavelengths and obtain the corresponding images for analysis [4-7]. Diffraction patterns recorded at multiple wavelengths have also been used for the lensless phase retrieval process [8-12]. In this case, the sample needs to be transparent and the intensity profiles are assumed to be identical at different wavelengths. It is not suitable for imaging stained samples such as histology slides.

A key aspect for on-chip multispectral microscopy is how to encode the spectral information into the intensity measurements. Current solutions either use filters to select the specific spectral information or sequentially illuminate the sample with different wavelengths to record the spectral information. In this letter, we discuss a new strategy to encode the spectral information using angle-tilted, wavelength-multiplexed ptychographic modulation. Our approach integrates 4 innovations for high-resolution multispectral on-chip imaging: 1) wavelength-multiplexed ptychographic phase retrieval [13-15], 2) slightly tilted illumination angles for breaking the ambiguities in mixed state reconstruction [16], 3) super-resolved up-sampling model for bypassing the pixel size limit of the image sensor [17], and 4) near-field ptychographic modulation for on-chip, large field of view imaging [18-20]. High-quality multi-color detection remains a challenge for on-chip microscopy due to the coherent artifacts in the conventional in-line configuration. Achieving high image quality and high resolution in lensless multispectral microscopy is significant for digital pathology where stained slides are examined over a large field of view.

Figure 1 shows the operation of the reported approach. We couple coherent lights from 5 laser diodes into a single-mode fiber and use it as a wavelength-multiplexed coherent source for sample illumination. The illumination beam, thus, contains 5 distinctive wavelengths from visible to near-infrared region: 415 nm, 488 nm, 532 nm, 660 nm, and 785 nm. We also add a prism at the illumination path for spectral dispersion. As such, lightwaves at different wavelengths hit the specimen at slightly different incident angles. At the detection path, we place a thin diffuser in-between the specimen and the monochromatic image sensor for encoding the spectral information into monochromatic measurements. Each acquired image represents an incoherent summation of 5 diffraction patterns at the 5 different wavelengths. By blindly scanning the sample (or the diffuser) to different x-y positions, we acquire a sequence of monochromatic images for reconstructing the 5 complex object profiles at the 5 wavelengths. The choice of sample or diffuser scanning is application driven. Both options work fine with solid samples and diffuser scanning is preferred for cell culture samples. In our implementation, we use a 12-megapixel Sony IMX226 monochromatic image sensor for image acquisition. The pixel size is 1.85 μm and the imaging field of view is 7.4 mm by 5.6 mm. The separation between the specimen and the image sensor is about 1 mm. The scanning step size is typically 1-2 μm.

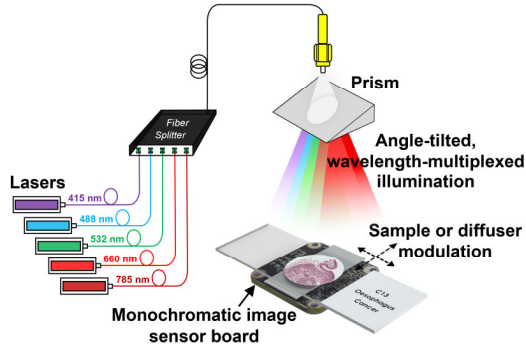

Fig. 1. Angle-tilted, wavelength-multiplexed ptychographic modulation for multispectral lensless on-chip microscopy.

The forward imaging model of our platform can be expressed as
$$I_i(x,y) = \sum_{t=1}^{5} |O_t(x,y) * PSF_{t,free}(d_1) \cdot P_t(x-x_i, y-y_i) * PSF_{t,free}(d_2) * PSF_{pixel}|^2_{\downarrow M}, \quad (1)$$
where $I_i(x,y)$ denotes the $i^{th}$ intensity measurement ($i = 1,2,...,I$), $O_t(x,y)$ represents the complex object exit wavefront at the $t^{th}$ wavelength, $P_t(x,y)$ represents the complex diffuser modulation pattern at the $t^{th}$ wavelength, $(x_i, y_i)$ denotes the $i^{th}$ positional shift of the diffuser, '·' stands for point-wise multiplication, '*' denotes the convolution operation, $d_1$ is the distance between the object and the diffuser, and $d_2$ is the distance between the diffuser and the image sensor. In Eq. (1), we use $PSF_{t,free}(d)$ to model the point spread function (PSF) for free-space propagation over distance $d$ at the $t^{th}$ wavelength (angular spectrum approach is used in our implementation). We use $PSF_{pixel}$ to model the PSF of the pixel response and assume it to be an averaging filter (all-ones matrix). Due to the relatively large pixel size (1.85 μm) of the imager, the captured image is an under-sampled version of the diffraction pattern. We use '$\downarrow M$' in the subscript of Eq. (1) to represent the down-sampling process ($M$=4 in our case). Based on all captured images $I_i(x,y)$ with the diffuser (or the sample) scanned to different lateral positions $(x_i, y_i)$s, we aim to recover the complex object $O_t(x,y)$ at different wavelengths.

The successful reconstruction of the multispectral object profiles relies on a good initial guess of the diffuser modulation patterns at different wavelengths. We perform a calibration experiment by sequentially turning on individual laser for sample illumination. For each wavelength, we move the sample to 300 different positions and recover both the sample and the diffuser profiles. The recovered diffuser patterns are then used as the initial guess for all subsequent experiments. We also note that, the multiplication of the tilted illumination waves with the objects is not modeled in Eq. (1). The effect of this operation is equivalent to shifting the diffuser pattern $P_t$ to different positions for $t$ = 1, 2, …, 5. These shifted modulation patterns have been recovered in the calibration experiment.

**Algorithm outline**

**Input**: Captured incoherent mixture of different wavelengths: $I_i$ ($i = 1,2,\cdots I$)
**Output**: Super-resolved object $O_t(x,y)$ and the diffuser $P_t(x,y)$, (t = 1,2,$\cdots$, 5)

1  Calculate the translational shift $(x_i, y_i)$ of the diffuser via cross-correlation
2  Initialize the $O_t(x,y)$. $P_t(x,y)$ is initialized in a calibration experiment
3  **for** $t$ =1:5 (different wavelengths)
4      Propagate to the diffuser plane: $O_{t,d_1}(x,y) = O_t(x,y) * PSF_{t,free}(d_1)$
5  end
6  **for** $n$ = 1:$N$ (different iterations)
7      **for** $i$ = 1:$I$ (different captured images)
8          **for** $t$ =1:5 (different wavelengths)
9              Shift the diffuser to $(x_i, y_i)$ position: $P_{t,i}(x,y) = P_t(x-x_i, y-y_i)$
10             Obtain the wavefront exiting the diffuser:
                  $\varphi_{t,i}(x,y) = O_{t,d_1}(x,y) \cdot P_{t,i}(x,y)$
11             Propagate to the detector plane
                  $\psi_{t,i}(x,y) = \varphi_{t,i}(x,y) * PSF_{t,free}(d_2)$
12             Obtain the target image $I_{t,i}(x,y)$ at the detector plane
                  $I_{t,i}(x,y) = |\psi_{t,i}(x,y)|^2_{\downarrow M}$
13         end
14         Generate the incoherent mixture: $I_{incoherent,i}(x,y) = \sum_{t=1}^{T=5} I_{t,i}(x,y)$
15         **for** $t$ = 1:5 (different wavelengths)
16             Update $\psi_{t,i}(x,y)$ using Eq. (2)
17             $\varphi'_{t,i}(x,y) = PSF_{t,free}(-d_2) * \psi'_{t,i}(x,y)$
18             Update the object $O_{t,d_1}(x,y)$ using Eq. (3)
19             Update $P_{t,i}(x,y)$ using Eq. (4)
20             Shift back the diffuser: $P_t(x,y) = P^{update}_{t,i}(x+x_i, y+y_i)$
21         end
22     end
23 end
24 **for** $t$ =1:5 (different wavelengths)
25     Propagate to the object plane: $O_t(x,y) = O^{update}_{t,d_1}(x,y) * PSF_{t,free}(-d_1)$
26 end

Fig. 2. The recovery process of the reported approach.

Figure 2 shows the recovery process for the reported approach. We first recover the positional shift of the diffuser (or the sample) via cross-correlation of the captured images. The object initial guess is obtained by up-sampling of the average of all measurements. In the iterative phase retrieval process, we propagate the $t^{th}$ object profile to the diffuser plane and obtain the exit wave $\varphi_{t,d_1}(x,y)$ in line 10. We then propagate the exit wave to the detector plane and obtain $\psi_{t,i}(x,y)$ in line 11. After repeating these steps for all five wavelengths, the intensity components $I_{t,i}(x,y)$ corresponding to different wavelengths are summed up to generate the incoherent mixture $I_{incoherent,i}(x,y)$ in line 14. We update $\psi_{t,i}(x,y)$ using the ratio between the actual measurement $I_i(x,y)$ and $I_{incoherent,i}(x,y)$, while keeping the phase unchanged:
$$\psi'_{t,i}(x,y) = \psi_{t,i}(x,y) \left( \frac{\sqrt{I_i(x,y)_{\uparrow M}}}{\sqrt{I_{incoherent,i}(x,y) * ones(M,M)_{\downarrow M \uparrow M}}} \right) \quad (2)$$

The updating process of Eq. (2) is motivated by the mixed-state formulation of multiplexed ptychography [13-15]. Here we introduce slightly different incident angles for lights at different wavelengths. The resulting diffuser modulation patterns, thus, become uncorrelated for different wavelengths, breaking the ambiguities in mixed state reconstruction [16]. We also note that, the image sizes of $\psi_{t,i}(x, y)$ and $I_i(x, y)$ are different in Eq. (2). If $I_i$ has a size of 100 by 100 pixels, $\psi_{t,i}$ will have 400 by 400 pixels, with an up-sampling factor $M=4$. The term '$I_i(x, y)_{\uparrow M}$' represents the nearest neighbor up sampling of the captured image $I_i$. In the denominator of Eq. (2), we first convolute $I_{incoherent,i}(x, y)$ with an averaging filter, i.e., $M$ by $M$ all-one matrix. We then perform $M$-times down-sampling followed by $M$-times nearest-neighbor up-sampling. This updating process enforces the intensity summation of every $M$ by $M$ small pixels equals the corresponding 1.85-µm large pixel in the captured image [17]. With the updated $\psi'_{t,i}(x, y)$, we update the object and the diffuser profiles corresponding to the $l^{th}$ wavelength via Eqs. (3) and (4) [21, 22]:

$$O^{update}_{t,d_1}(x, y) = O_{t,d_1}(x, y) + \frac{conj(P_{t,i}(x,y)) \cdot \{\varphi'_{t,i}(x,y) - \varphi_{t,i}(x,y)\}}{(1-\alpha_p)|P_{t,i}(x,y)|^2 + \alpha_p|P_{t,i}(x,y)|^2_{max}}, \quad (3)$$

$$P^{update}_{t,i}(x, y) = P_{t,i}(x, y) + \frac{conj(O_{t,d_1}(x,y)) \cdot \{\varphi'_{t,i}(x,y) - \varphi_{t,i}(x,y)\}}{(1-\alpha_{obj})|O_{t,d_1}(x,y)|^2 + \alpha_{obj}|O_{t,d_1}(x,y)|^2_{max}}, \quad (4)$$

where 'conj' denotes conjugate, and $\alpha_{obj}$ and $\alpha_p$ are algorithm weights in the iterative process (we set them to 0.9). The recovery process typically converges within 5 iterations in experiments reported in this work. We have performed a simulation study to validate the reported imaging concept in a supplementary note [23]. In our experiments, we acquire 1000 raw images with an acquisition time of ~40 seconds. The processing time for 1000 images with 1024 pixels by 1024 pixels each is ~13 mins using a Dell XPS Tower computer.

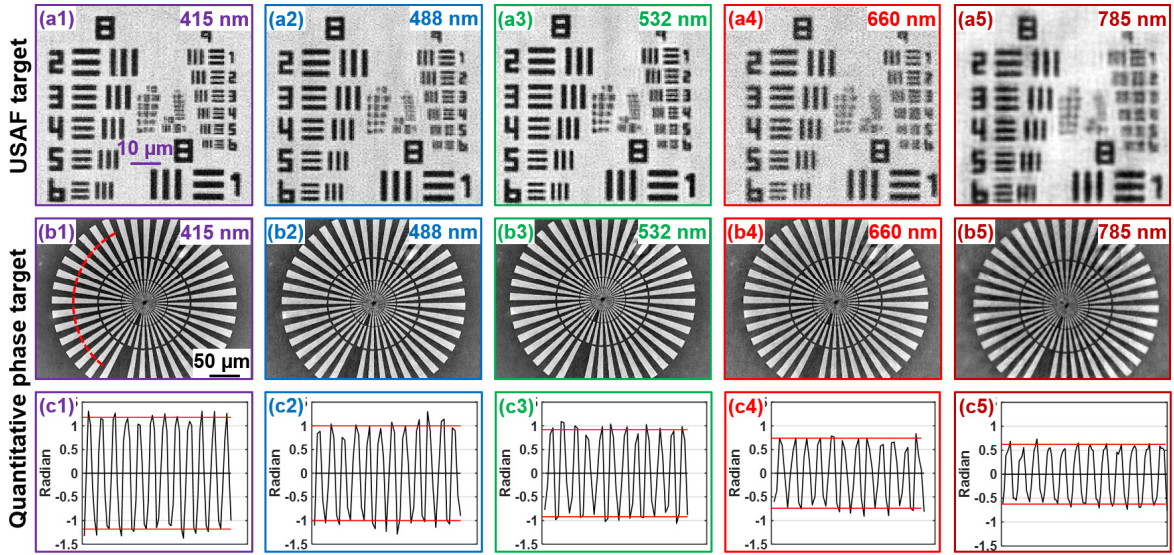

Fig. 3. Validating the reported approach using a USAF resolution target (a) and a phase target (b). 0.55-µm linewidth from group 9, element 6 can be clearly resolved in (a1). (c) Line trace of phase across the phase target. The two red lines in (c) indicates the ground-truth phase variations of the target. The ultimate resolution is limited by the maximum scattering angle that can be detected by the imager, not the pixel size.

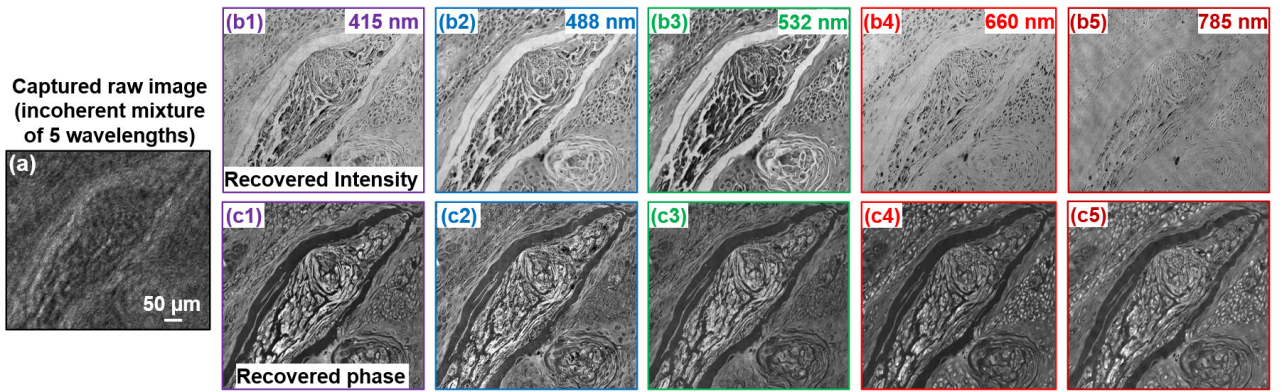

Fig. 4. (a) The captured raw image of a stained tissue section (1000 raw images in total). The recovered intensity (b) and phase (c) at the 5 wavelengths.

We first validate our platform using a USAF resolution target and a quantitative phase target in Fig. 3. Figure 3(a) shows the recovered resolution target at the 5 wavelengths, where we can resolve the 0.55-μm linewidth from group 9, element 6 at the 415-nm wavelength. Figure 3(b) shows the recovered quantitative phase using the reported platform and the line trace is plotted in Fig. 3(c). We conclude that the recovered phase is in a good agreement with the ground-truth of the phase target (two red lines in Fig. 3(c)), validating the quantitative nature of the reported approach.

In the second experiment, we test our platform using a hematoxylin and eosin-stained esophagus cancer slide. Figure 4(a) shows the captured raw image, which is an incoherent mixture of diffraction patterns at 5 wavelengths. Figure 4(b)-(c) show the recovered intensity and phase at the 5 wavelengths.

In the third experiment, we generate color images of two stained histology slides from our multiplexed reconstructions and compare them with two other approaches. Figure 5(a1) shows our multiplexed reconstruction of a skin cancer immunohistochemistry slide labeled with Ki-67 markers. Figure 5(a2) shows our multiplexed reconstruction of the hematoxylin and eosin-stained esophagus cancer slide (same sample as that in Fig. 4). For Fig. 5(a), we generate the color images by combining the reconstructions at 488 nm, 532 nm, and 660 nm. As a comparison, Fig. 5(b) shows the color image generated by sequential illumination of the three wavelengths with no multiplexing. Figure 5(c) shows the color images captured using an upright Nikon microscope with a 20X, 0.75 numerical aperture (NA) objective lens. The slight color difference is due to the use of a broadband light source in the microscope.

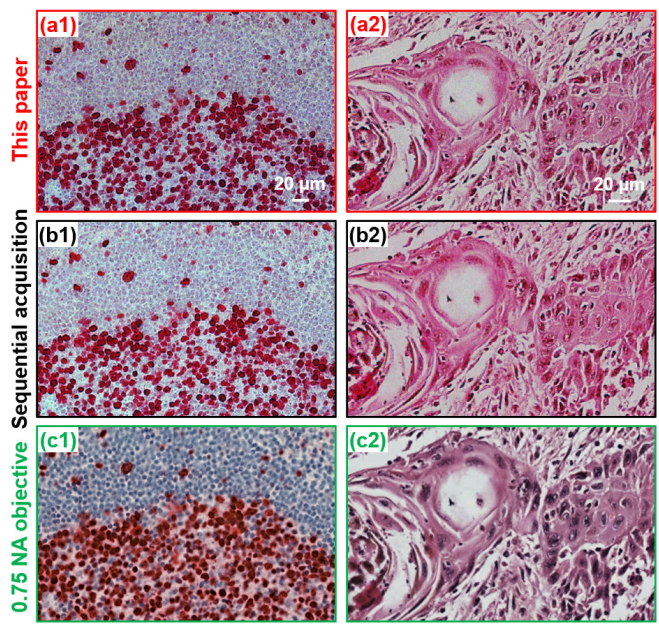

Fig. 5. Comparison between the reported approach (a), sequential acquisition for color imaging (b), and images captured by a 0.75 NA objective (c).

In summary, we report an angle-tilted, wavelength-multiplexed ptychographic modulation approach for multispectral lensless on-chip microscopy. By introducing slightly different incident angles for illumination waves at different wavelengths, the diffuser modulation patterns become uncorrelated with each other, thus breaking the ambiguities in mixed state ptychographic reconstruction. By integrating an up-sampling procedure in the multiplexed phase retrieval process, we demonstrate a half-pitch resolution of 0.55 μm using an imager with 1.85-μm pixel size. This achieved resolution is the highest among the reported optical ptychographic implementations. Given this is a lensless platform, imaging performance can be maintained over the entire field of view without aberration issues. It may have advantages compared to other lens-based approaches such as Fourier ptychographic microscopy [24]. One future direction for the reported approach is to perform hyperspectral imaging using a supercontinuum laser. Another direction for the reported approach is to employ LEDs for sample illumination, where we can better model the spectral bandwidth of the LED emission.

**Funding.** National Science Foundation 1510077 and 1700941.

**Disclosures.** The authors declare no conflicts of interest.

**References**
1. F. Ghaznavi, A. Evans, A. Madabhushi, and M. Feldman, Annual Review of Pathology: Mechanisms of Disease 8, 331-359 (2013).
2. A. Greenbaum, A. Feizi, N. Akbari, and A. Ozcan, Opt. Express 21, 12469-12483 (2013).
3. Y. Wu, Y. Zhang, W. Luo, and A. Ozcan, Scientific Reports 6, 28601 (2016).
4. S. A. Lee, R. Leitao, G. Zheng, S. Yang, A. Rodriguez, and C. Yang, PLOS ONE 6, e26127 (2011).
5. G. Zheng, S. A. Lee, Y. Antebi, M. B. Elowitz, and C. Yang, Proceedings of the National Academy of Sciences 108, 16889-16894 (2011).
6. Y. Bian, Q. Liu, Z. Zhang, D. Liu, A. Hussian, C. Kuang, H. Li, and X. Liu, Optics and Lasers in Engineering 111, 25-33 (2018).
7. S. A. Lee, J. Erath, G. Zheng, X. Ou, P. Willems, D. Eichinger, A. Rodriguez, and C. Yang, PLOS ONE 9, e89712 (2014).
8. E. Mathieu, C. D. Paul, R. Stahl, G. Vanmeerbeeck, V. Reumers, C. Liu, K. Konstantopoulos, and L. Lagae, Lab on a Chip 16, 3304-3316 (2016).
9. P. Bao, F. Zhang, G. Pedrini, and W. Osten, Optics Letters 33, 309-311 (2008).
10. C. Zuo, J. Sun, J. Zhang, Y. Hu, and Q. Chen, Optics express 23, 14314-14328 (2015).
11. C. Allier, S. Morel, R. Vincent, L. Ghenim, F. Navarro, M. Menneteau, T. Bordy, L. Hervé, O. Cioni, and X. Gidrol, Cytometry Part A 91, 433-442 (2017).
12. L. Herve, O. Cioni, P. Blandin, F. Navarro, M. Menneteau, T. Bordy, S. Morales, and C. Allier, Biomed. Opt. Express 9, 5828-5836 (2018).
13. P. Thibault and A. Menzel, Nature 494, 68 (2013).
14. D. J. Batey, D. Claus, and J. M. Rodenburg, Ultramicroscopy 138, 13-21 (2014).
15. S. Dong, R. Shiradkar, P. Nanda, and G. Zheng, Biomed. Opt. Express 5, 1757-1767 (2014).
16. P. Li, T. Edo, D. Batey, J. Rodenburg, and A. Maiden, Opt. Express 24, 9038-9052 (2016).
17. D. Batey, T. Edo, C. Rau, U. Wagner, Z. Pešić, T. Waigh, and J. Rodenburg, Physical Review A 89, 043812 (2014).
18. M. Stockmar, P. Cloetens, I. Zanette, B. Enders, M. Dierolf, F. Pfeiffer, and P. Thibault, Scientific reports 3, 1927 (2013).
19. S. Jiang, J. Zhu, P. Song, C. Guo, Z. Bian, R. Wang, Y. Huang, S. Wang, H. Zhang, and G. Zheng, Lab on a Chip 20, 1058-1065 (2020).
20. F. Zhang and J. Rodenburg, Physical Review B 82, 121104 (2010).
21. A. Maiden, D. Johnson, and P. Li, Optica 4, 736-745 (2017).
22. P. Song, S. Jiang, H. Zhang, Z. Bian, C. Guo, K. Hoshino, and G. Zheng, Opt. Lett. 44, 3645-3648 (2019).
23. https://figshare.com/articles/Supplementary_Note/12307490.
24. G. Zheng, R. Horstmeyer, and C. Yang, Nature photonics 7, 739 (2013).